\documentclass[intlimits,twoside,a4paper]{article}

\usepackage[cp1251]{inputenc}
\usepackage[eqsecnum]{cmpj3}
\usepackage{multirow}
\usepackage{multicol}
\usepackage{hyperref}

\issue{2022}{25}{4}{43704}
\doinumber{10.5488/CMP.25.43704}

\title[Temperature dependence of dielectric permittivity in incommensurately modulated phase of ammonium fluoroberyllate]
{Temperature dependence of dielectric permittivity in incommensurately modulated phase of ammonium fluoroberyllate}

\author[B. I. Horon, O. S. Kushnir, P. A. Shchepanskyi, V. Yo. Stadnyk]{B. I. Horon\orcid{0000-0002-4595-5789}\refaddr{label1, label2}\thanks{Corresponding author: \email{bohdan.horon@lnu.edu.ua}}, O. S. Kushnir\orcid{0000-0002-1545-7666}\refaddr{label2}, P. A. Shchepanskyi\orcid{0000-0003-0841-7508}\refaddr{label1}, V. Yo. Stadnyk\orcid{0000-0001-7820-7886}\refaddr{label1}}
\addresses{
\addr{label1} General Physics Department, Ivan Franko National University of Lviv, 23 Drahomanov Street, 79005 Lviv, Ukraine
\addr{label2} Optoelectronics and Information Technologies Department, Ivan Franko National University of Lviv, \\107 gen. Tarnavskyi Street, 79013 Lviv, Ukraine}


\date{Received July 15, 2022, in final form September 28, 2022}

\begin{document}

\maketitle

\begin{abstract}
    We study the temperature dependence of dielectric permittivity along the polar axis for ferroelectric ammonium fluoroberyllate (AFB) crystal in the vicinity of its phase transition points.
    The experimental data within incommensurately modulated phase of AFB is compared with the predictions of phenomenological models known from the literature: the Curie-Weiss (CW) law, the generalized Curie-Weiss (GCW) law, and the models by Le\-va\-nyuk and Sannikov (LS) and by Prelov\v{s}ek, Levstik and Filipi\v{c} (PLF) suggested for improper ferroelectrics.
    It is shown that the LS approach describes  the temperature behavior of the dielectric permittivity for the AFB crystal better than the CW, GWC and PLF models.
    The main physical reasons of this situation are elucidated.
 
	\printkeywords    
\end{abstract}

\section{Introduction} 

Ammonium fluoroberyllate (NH$_4$)$_2$BeF$_4$ (or AFB) is an improper ferroelectric crystal that belongs to a large A$_2$BX$_4$ family.
It undergoes two phase transitions (PTs) approximately at the temperatures $T_{\textrm{C}} \approx 177$~K and $T_{\textrm{i}} \approx 183$~K \cite{strukovetal1973, gesiozawa1974, levanyuksannikov1976, strukovarutyunovauesu1982, prelovseklevstikfilipic1983, srivastavekloosterkoetzle1999}, which separate a low-temperature ferroelectric phase, an intermediate incommensurate phase and a high-temperature paraelectric phase.
Although the AFB crystals have been thoroughly studied during decades (see, e.g.,  \cite{srivastavekloosterkoetzle1999, strukovsmirnov1986, palatinussmaalen2004, brikkityk2007, strukovlevanyuk1998, palatinusamamismaalen2004}), some problems of their PTs and critical phenomena still remain a matter of dispute.

In particular, AFB reveals an intriguing temperature dependence of its dielectric permittivity: unlike the optical birefringence and many other characteristics, dielectric anomaly at $T_\textrm{i}$ is in fact absent, while the $T_\textrm{C}$ point is marked by only a weak dielectric peak \cite{gesiozawa1974, levanyuksannikov1976, strukovarutyunovauesu1982, prelovseklevstikfilipic1983, jakubasczpala1984}.
The dielectric properties of AFB have been the main subject of theoretical studies by  Levanyuk and  Sannikov \cite{levanyuksannikov1976} and by  Prelov\v{s}ek,  Levstik and  Filipi\v{c} \cite{prelovseklevstikfilipic1983} (abbreviated respectively as LS and PLF), which are both based upon the hypothesis of improper ferroelectricity in AFB.
In spite of this fact, the final expressions obtained in~\cite{levanyuksannikov1976, prelovseklevstikfilipic1983} turn out to be different in many respects.

The other notable fact is that there has been no study where an experimental temperature dependence of the dielectric permittivity for the AFB crystals would be simultaneously compared with different theoretical formulae in order to estimate the advantages and shortcomings of the latter.
The only exception, our recent work \cite{horonkushnirstadnykkashuba2020}, represents a short technical report based upon contemporary methods of nonlinear fitting and statistical techniques (see the works \cite{kushnirshopavlokh2008, girnykklymovychkushnirshopa2014}).
Although a number of weak methodical points associated with fitting \cite{goldsteinmorrisyen2004, bauke2007, perline2005} are omitted in this work, no physical reasoning and data interpretation have been made there.

In the present study we compare all of the available phenomenological approaches which can, in principle, be applied to describe the dielectric properties of the AFB crystals and explain why the LS theory \cite{levanyuksannikov1976} exceeds the performance of other approaches and perfectly fits   the experimental data for dielectric permittivity.

\section{Experimental data and short description of theoretical models} 
\label{sec:exp_models}

\begin{figure}[!t]
	\begin{minipage}[h]{0.49\textwidth}
		\center{\includegraphics[width=0.85\textwidth]{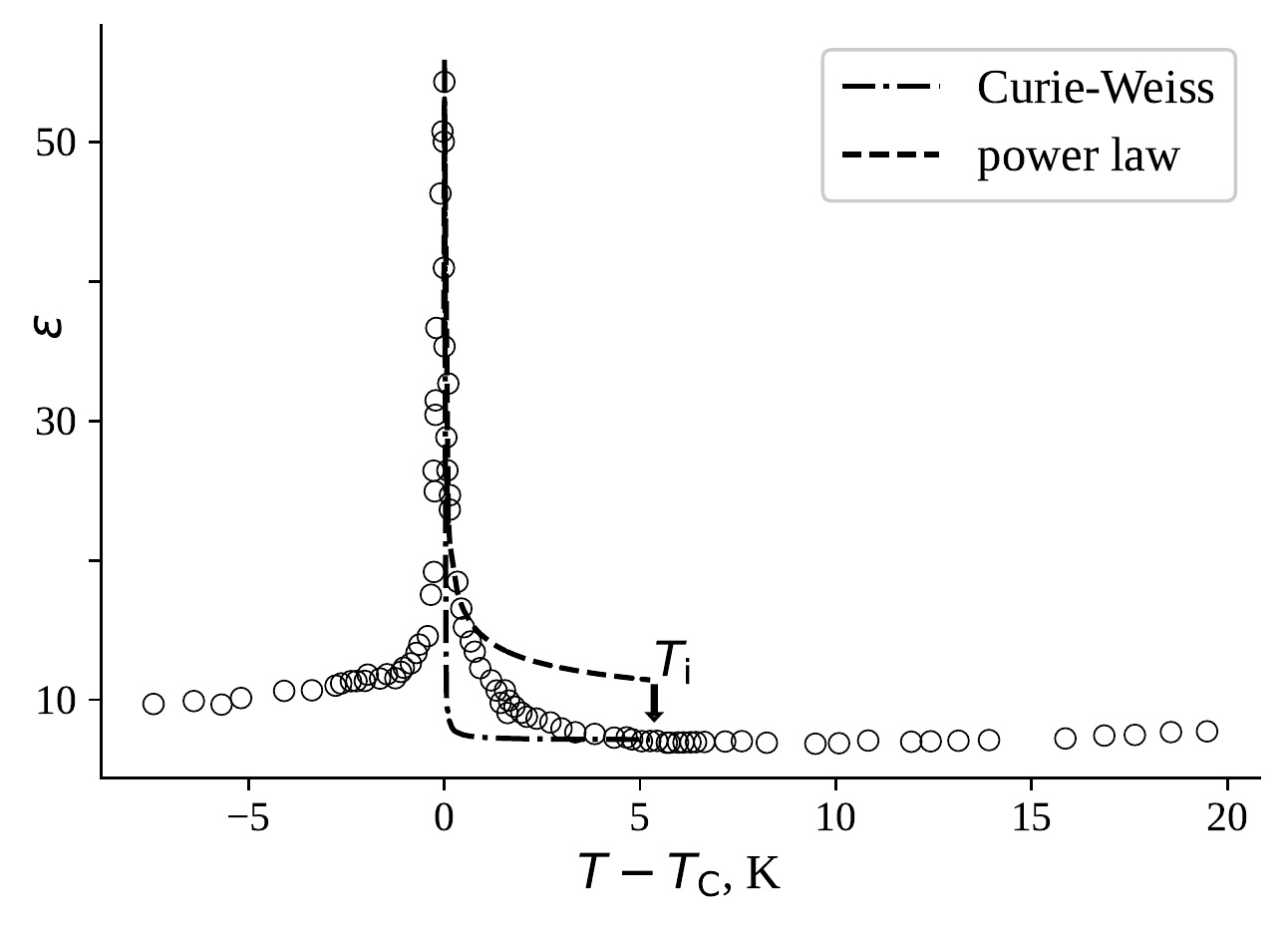} \\ a}
	\end{minipage}
	\hfill
	\begin{minipage}[h]{0.49\linewidth}
		\center{\includegraphics[width=0.85\textwidth]{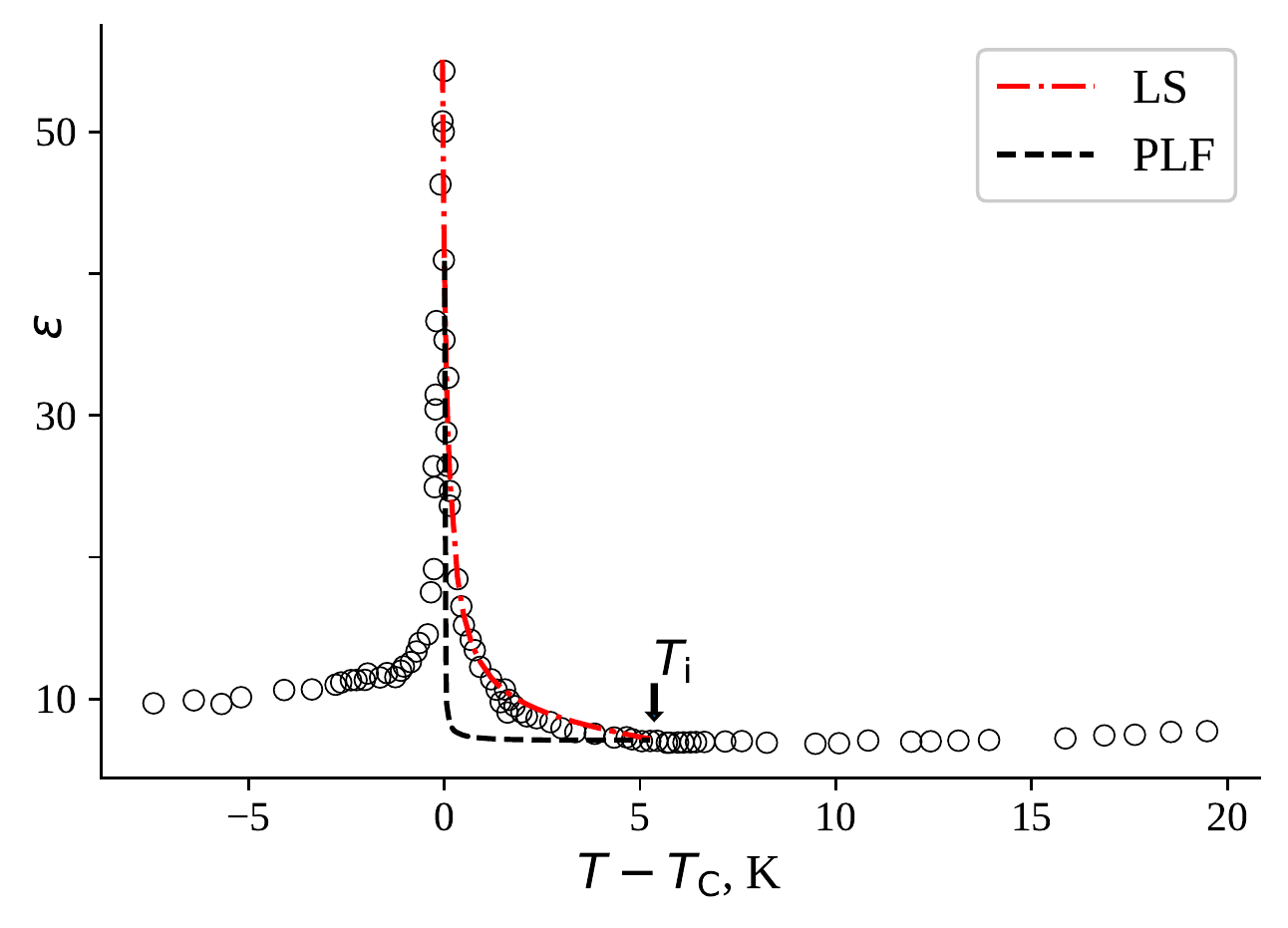} \\ b}
	\end{minipage}
	\caption{(Colour online) Experimental temperature dependence $\varepsilon(T)$ of dielectric permittivity for the AFB crystals (circles) and its fitting (lines) with the theoretical models (1), (2) (panel a) and (3), (4) (panel b) within the incommensurate phase.}
	\label{fig:model_comp}
\end{figure}

A single crystal of AFB for our studies was grown from aqueous solution of a stoichiometric mixture of NH$_4$ and BeF$_2$, using a standard method of slow cooling.
The dielectric permittivity was measured along the polar axis with an automated capacitive apparatus (the temperature region 170--200~K, the tolerance of temperature measurement $\sim$0.1~K, and the working frequency 1~kHz).
Figure~\ref{fig:model_comp} displays the experimental temperature dependence of the dielectric permittivity for the AFB crystals.
As seen from figure~\ref{fig:model_comp}, no anomaly is visible at $T_{\textrm{i}}$, in compliance with the main bulk of experimental data known from the literature.

Note also that the maximum dielectric permittivity detected by us ($\varepsilon_\textrm{max} \approx 55$) correlates well with the data obtained in the earlier measurements for the improper AFB crystals ($\varepsilon_\textrm{max} \approx$ 35--160 \cite{strukovetal1973, gesiozawa1974, levanyuksannikov1976, strukovarutyunovauesu1982, prelovseklevstikfilipic1983, jakubasczpala1984, hoshinoetal1958}).
This is contrary to the proper ferroelectrics where the values $\varepsilon_\textrm{max} \sim 10^3-10^5$ are often detected (see~\cite{strukovlevanyuk1998}).
Such a small $\varepsilon_{\textrm{max}}$ peak can indeed be successfully interpreted using the idea that the dielectric anomaly in improper ferroelectrics is a secondary effect while a true order parameter has the symmetry different from that of spontaneous electric polarization. 
For the same or somewhat different reasons, weak dielectric anomalies are also typical of ferroelastics \cite{kushnirshopapolovynko2007}, multiferroics \cite{kundysetal2010} and ferroelectrics with noticeable amounts of structural defects \cite{kushnirshopavlokh2008, girnykklymovychkushnirshopa2014, otkoetal1993}.
 
Now we proceed to phenomenological consideration of the dielectric properties of AFB.
Since the both LS \cite{levanyuksannikov1976} and PLF \cite{prelovseklevstikfilipic1983} models have not dealt with the $\varepsilon(T)$ function within the ferroelectric phase, we  analyze the dielectric data only within the incommensurate and paraelectric phases.
Another important point is a so-called `background' dielectric permittivity $\varepsilon_{\textrm{b}}$ which can be independent of the PTs.
The problem of the background versus the PT-driven anomaly is familiar in examining the specific heat of ferroics, since the appropriate anomaly is often comparable with the lattice contributions (see, e.g.,  \cite{uetanietal1998, matsuoetal2003}).
However, this is not so in the field of dielectric studies of proper ferroelectrics where huge anomalous peaks are mostly observed, so that neglecting  the background would not hinder the possibility of obtaining highly accurate fitting data.
As a consequence, even a constant background term $\varepsilon_{\textrm{b}}$ has rarely been considered in the dielectric studies of ferroelectrics, not to mention a temperature dependent background $\varepsilon_{\textrm{b}}(T)$.
Still, a few relevant exceptions are known from the literature \cite{kushnirshopavlokh2008, sandvoldcourtens1983}.
Of course, consideration of $\varepsilon_{\textrm{b}}$ can become very important when improper ferroelectrics like AFB are addressed.
Finally, the both LS and PLF models \cite{levanyuksannikov1976}\cite{prelovseklevstikfilipic1983} treat the dielectric function as temperature-independent within the paraelectric phase and this is consistent with  our both experimental results and the whole bulk of the literature data (especially with that obtained in a broad temperature range \cite{kushnirshopapolovynko2007}).
Therefore, we restrict ourselves to the simplest assumption $\varepsilon_{\textrm{b}} = \textrm{const}$.

Next, the dielectric permittivity within the incommensurate phase can, in principle, be described by one of the following theoretical models.

\textbf{Model (1)}. A canonical Curie-Weiss law with a constant background $\varepsilon_{\textrm{b}}$ and a Curie constant $C_{\textrm{CW}}$:
\begin{equation} \label{curieweiss}
    \varepsilon(T) = \varepsilon_{\textrm{b}} + \frac{C_{\textrm{CW}}}{T - T_\textrm{C}}.
\end{equation}

\textbf{Model (2)}. A power law representing generalization of the Curie-Weiss formula~(\ref{curieweiss}), with the exponent $\gamma > 1$:
\begin{equation} \label{genpower}
    \varepsilon(T) = \varepsilon_{\textrm{b}} + \frac{C_{\gamma}}{(T - T_\textrm{C})^{\gamma}}.
\end{equation}

\textbf{Model (3)}. The LS model \cite{levanyuksannikov1976} for the incommensurate phase which in fact states that 
\begin{equation} \label{lsmodel}
    \varepsilon(T) = \varepsilon_{\textrm{b}} + \varepsilon_{\textrm{b}}^2 A\, t \frac{6 + t}{4 - t},
\end{equation}
where $A$ is a constant and $t$ implies the reduced temperature:
\begin{equation*}
    t = \frac{T_{\textrm{i}} - T}{T_{\textrm{i}} - \theta},
\end{equation*}
with $\theta$ ($T_{\textrm{C}} < \theta < T_{\textrm{i}}$) being an instability point for the order parameter. 
It can be defined in terms of the distance $\Delta T$ from the $T_{\textrm{C}}$ point ($\theta = T_{\textrm{C}} + \Delta T$). Note that $\Delta T$ can be expressed in terms of the free-energy expansion \cite{levanyuksannikov1976} (see section \ref{sec:comparison}).

According to formula (\ref{lsmodel}), the $\varepsilon(T)$ function diverges at $t =$ 4 and tends to $\varepsilon = \varepsilon_{\textrm{b}}$ at $T = T_{\textrm{i}}$.
Since the model predicts the same constant value $\varepsilon = \varepsilon_{\textrm{b}}$ in the paraelectric phase, the dielectric function is continuous at the incommensurate--paraelectric PT, while the slope of $\varepsilon(T)$ curve suffers an abrupt change at $T_{\textrm{i}}$.
Although the authors \cite{levanyuksannikov1976} themselves have defined the applicability limits of formula (\ref{lsmodel}) as a narrow temperature region below the $T_{\textrm{i}}$ point (e.g., as a region of small $t$ in terms adopted in this work), we have checked the model (3) in the overall range between the temperatures $T_{\textrm{C}}$ and $T_{\textrm{i}}$.

\textbf{Model (4)}. The PLF model \cite{prelovseklevstikfilipic1983} for the incommensurate phase:
\begin{equation} \label{plfmodel}
    \varepsilon(T) = \varepsilon_{\textrm{b}} + \frac{\varepsilon_\textrm{b}}{c}\bigg\{\frac{E(\tau)}{(1 - \tau^2)K(\tau)} - 1\bigg\},
\end{equation}
where $c$ is a constant, while $K(\tau)$ and $E(\tau)$ denote the complete elliptic integrals of the first and second kinds, respectively.
The authors \cite{prelovseklevstikfilipic1983} have not linked the elliptic modulus $\tau$ with the PT parameters.
Nonetheless, the relation $\tau = \frac{T_{\textrm{i}} - T}{T_{\textrm{i}} - T_{\textrm{C}}}$ can be postulated due to the properties $\tau \rightarrow 0$ and $\tau \rightarrow 1$ holding respectively at $T \rightarrow T_{\textrm{i}}$ and $T \rightarrow T_{\textrm{C}}$ \cite{prelovseklevstikfilipic1983}.
According to the PLF approach \cite{prelovseklevstikfilipic1983}, formula (\ref{plfmodel}) can be assumed to be applicable within the entire incommensurate phase.
A criticality at $T_{\textrm{C}}$ in formula (\ref{plfmodel}) is due to the behaviour of the terms $(1 - \tau^2)$ and $K(\tau)$ at $\tau \rightarrow 1$.
Since the equality $K(\tau) = E(\tau)$ takes place at $\tau = 0$, we have $\varepsilon = \varepsilon_{\textrm{b}}$ at $T = T_{\textrm{i}}$.
Finally, the $T_{\textrm{i}}$ point, which corresponds to the anomaly found experimentally in the specific heat, is hardly detectable in the dielectric permittivity.
Similarly to the LS model, the only track of this PT is a change in the $\varepsilon(T)$ slope, which can be detected in a smoothed temperature dependence $\rd \varepsilon(T)/\rd T$ (not shown in figure~\ref{fig:model_comp}).

It is worthwhile that, contrary to the models (1) and (2), the temperature-independent background~$\varepsilon_{\textrm{b}}$ is introduced into the formulae (\ref{lsmodel}) and (\ref{plfmodel}) directly from the expansion of free energy $\Phi$ (in fact, via the relation $\Phi \sim P^2/(2\varepsilon_{\textrm{b}})$, with $P$ being the electric polarization \cite{levanyuksannikov1976} --- see section \ref{sec:comparison}).
Notice also that, in case of AFB, we have evidently different experimental background levels found in the paraelectric and ferroelectric phases (see figure~\ref{fig:model_comp}).
Finally, theoretical considerations \cite{levanyuksannikov1976} testify that it is unnecessary to retain any temperature-dependent $\varepsilon_{\textrm{b}}(T)$ terms.

\section{Fitting the results and their discussion} 

Now we  fit our experimental data $\varepsilon(T)$ using the phenomenological models (1)--(4), determine the best model and explain in detail the practical advantages and disadvantages of those models.
Procedures of nonlinear fitting are implemented according to a standard Levenberg--Marquardt algorithm.
A goodness-of-fit is evaluated with $\chi^2$ and Wald--Wolfowitz statistical tests.
Finally, the error margins for the model parameters are found with a bootstrap technique, using 2000 synthetic datasets. 
The appropriate details are elucidated elsewhere \cite{horonkushnirstadnykkashuba2020}. Figure~\ref{fig:model_comp} illustrates the fitting results and table~\ref{tab:model_comp} displays a short account of the main model parameters.

\begin{table}[htb]    
\caption{Some of fitting parameters of the $\varepsilon(T)$ dependence corresponding to phenomenological models~(1)--(4).} 
\label{tab:model_comp}
\vspace{0.5cm}
\begin{center}    
\begin{tabular}{|c||c|c|}    
\hline    
\multirow{2}{*}{Model} & \multicolumn{2}{|c|}{Results} \strut \\
    \cline{2-3}
    & Parameter & Value\strut\\    
\hline    
    (1)  & $C_{\textrm{CW}}$ & 0.179 \strut\\    
\hline    
\multirow{2}{*}{(2)} & $C_{\gamma}$ & 7.392  \strut\\    
\cline{2-3}    
         & $\gamma$ &  0.326 \strut\\    
\hline  
\multirow{2}{*}{(3)} & $A$ & 0.0094 \strut\\    
\cline{2-3}    
      & $\Delta T$,~K &  4.01 \strut\\    
\hline    
     (4) & $c$ & 30.373 \strut\\    
\hline    
\end{tabular}    
\end{center}    
\end{table}

The Curie-Weiss law underestimates the experimental $\varepsilon(T)$ curve at the temperatures more or less distant from the PT (figure~\ref{fig:model_comp}a) and so fails to appropriately fit  the dielectric permittivity.
Moreover, the Curie-Weiss fit reveals too large Z-score (see table~\ref{tab:stat_results}).
The PLF model (4) has the characteristics similar to those of the Curie-Weiss law (see figure~\ref{fig:model_comp}b and table~\ref{tab:stat_results}).
The other pattern takes place for the model~(2), which corresponds to generalized power-law for the temperature dependence $\varepsilon(T)$.
Here, the fitting function overestimates most of the experimental data point values and fails to catch the background (see figure~\ref{fig:model_comp}a), whereas the Z-score is just as large as that for the other models mentioned above.
Moreover, the model provides a $\gamma$ value noticeably less than unity (see table~\ref{tab:model_comp}), which is a physically shallow result. 
Hence, the generalized power law for the dielectric permittivity of AFB is also insufficient.

\begin{table}[htb]
    \centering
    \caption{Results of $\chi^2$ and Wald--Wolfowitz tests for phenomenological models (1)--(4).}
    \label{tab:stat_results}
    \vspace{0.5cm}
    \begin{tabular}{|c||c|c|}
    \hline
    \multirow{2}{*}{Model}  & \multicolumn{2}{|c|}{Statistical Tests} \strut \\
    \cline{2-3}
         & Parameter & Value \strut \\
    \hline
    \multirow{3}{*}{(1)} & $\chi^2$ & 5252.33 \strut \\
    \cline{2-3}
    & Reduced $\chi^2$ & 165.22 \strut \\
    \cline{2-3}
    & Z-score & $-3.45$ \strut \\
    \cline{2-3}
    \hline
    \multirow{4}{*}{(2)} & $\chi^2$ & 1525.25 \strut \\
    \cline{2-3}
    & Reduced $\chi^2$ & 47.66 \strut \\
    \cline{2-3}
    & Z-score & $-4.85$ \strut \\
    \cline{2-3}
    & Correlation($C_{\gamma}$, $\gamma$) & $-0.87$ \strut \\
    \hline
    \multirow{4}{*}{(3)} & $\chi^2$ & 320.87 \strut \\
    \cline{2-3}
    & Reduced $\chi^2$ & 9.72 \strut \\
    \cline{2-3}
    & Z-score & 0.34 \strut \\
    \cline{2-3}
    & Correlation($A$, $\Delta T$) & $-0.95$ \strut \\
    \hline
    \multirow{3}{*}{(4)} & $\chi^2$ & 2189.92 \strut \\
    \cline{2-3}
    & Reduced $\chi^2$ & 72.99 \strut \\
    \cline{2-3}
    & Z-score & $-4.39$ \strut \\
    \hline
    \end{tabular}
\end{table}

On the contrary, the theoretical curve referred to the LS model (3) fits fairly well the experimental data, and the appropriate statistical tests provide quite satisfactory results (see figure~\ref{fig:model_comp}b, table~\ref{tab:model_comp} and table~\ref{tab:stat_results}). 
Moreover, it becomes evident that the model (3) can in fact be applied in the entire temperature range under study, contrary to the cautions of the authors \cite{levanyuksannikov1976}.
Finally, the term $\varepsilon_{\textrm{b}} = 7.12$ found from the LS fitting (not shown in table~\ref{tab:model_comp}) turns out to be very close to the experimental dielectric background averaged over the paraelectric phase.
In other words, the LS phenomenology obviously exceeds the performance of the other models.
For completeness, we list the PT points derived with the model (3), which are not displayed in table~\ref{tab:model_comp} for the sake of brevity: $T_{\textrm{C}} = 177.64$~K (found from the dielectric peak), $T_{\textrm{i}} = 183.19$~K and $\theta = 181.65$~K.
Finally, the confidence intervals for the model parameters $A$ and $\Delta T$ are given respectively by $-$0.0053--0.0272 and 3.668--4.366 (cf. with the data of table~\ref{tab:model_comp}).

\begin{figure}[!t]
	\begin{minipage}[h]{0.49\textwidth}
		\center{\includegraphics[width=0.85\textwidth]{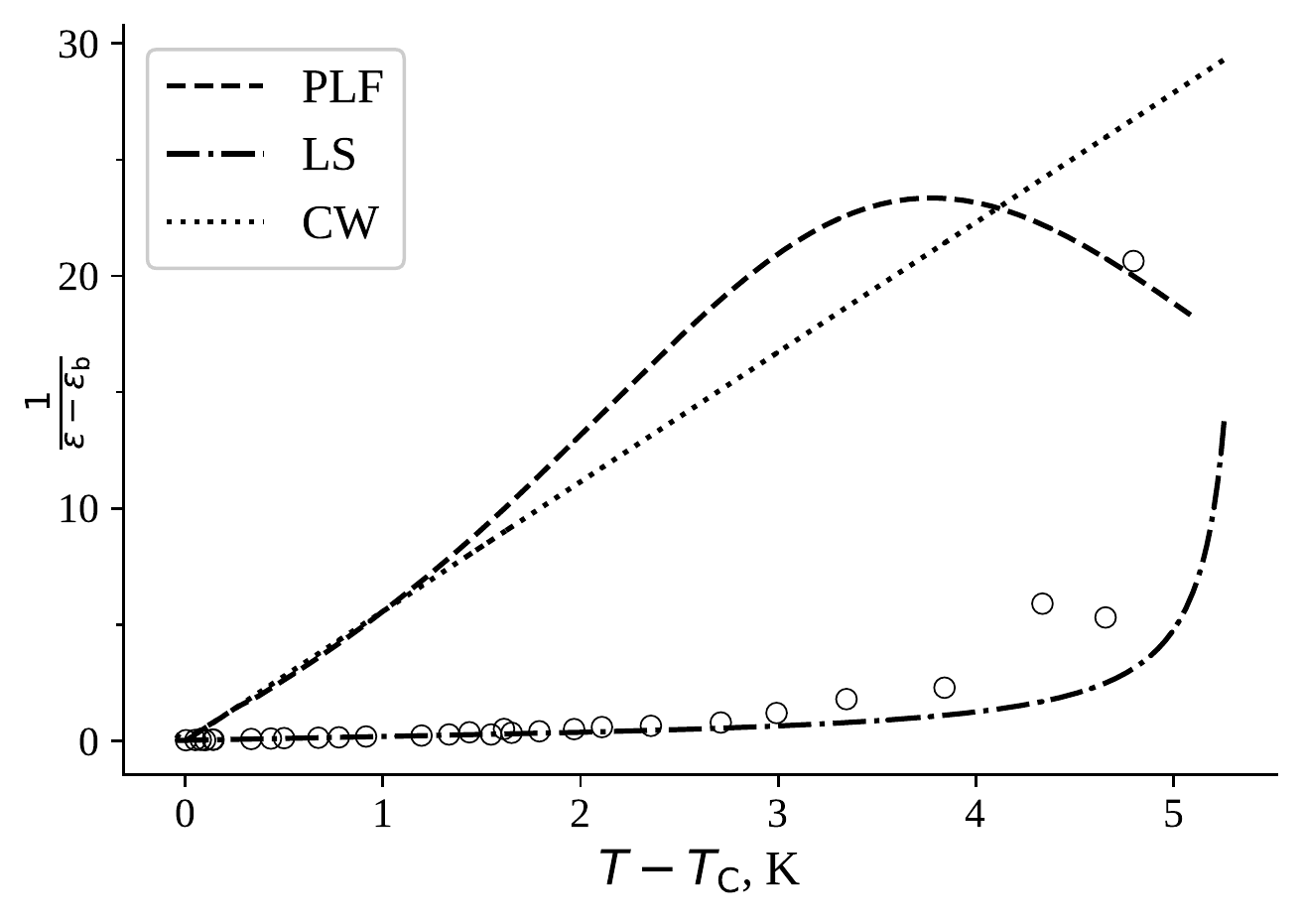} \\ a}
	\end{minipage}
	\hfill
	\begin{minipage}[h]{0.49\linewidth}
		\center{\includegraphics[width=0.85\textwidth]{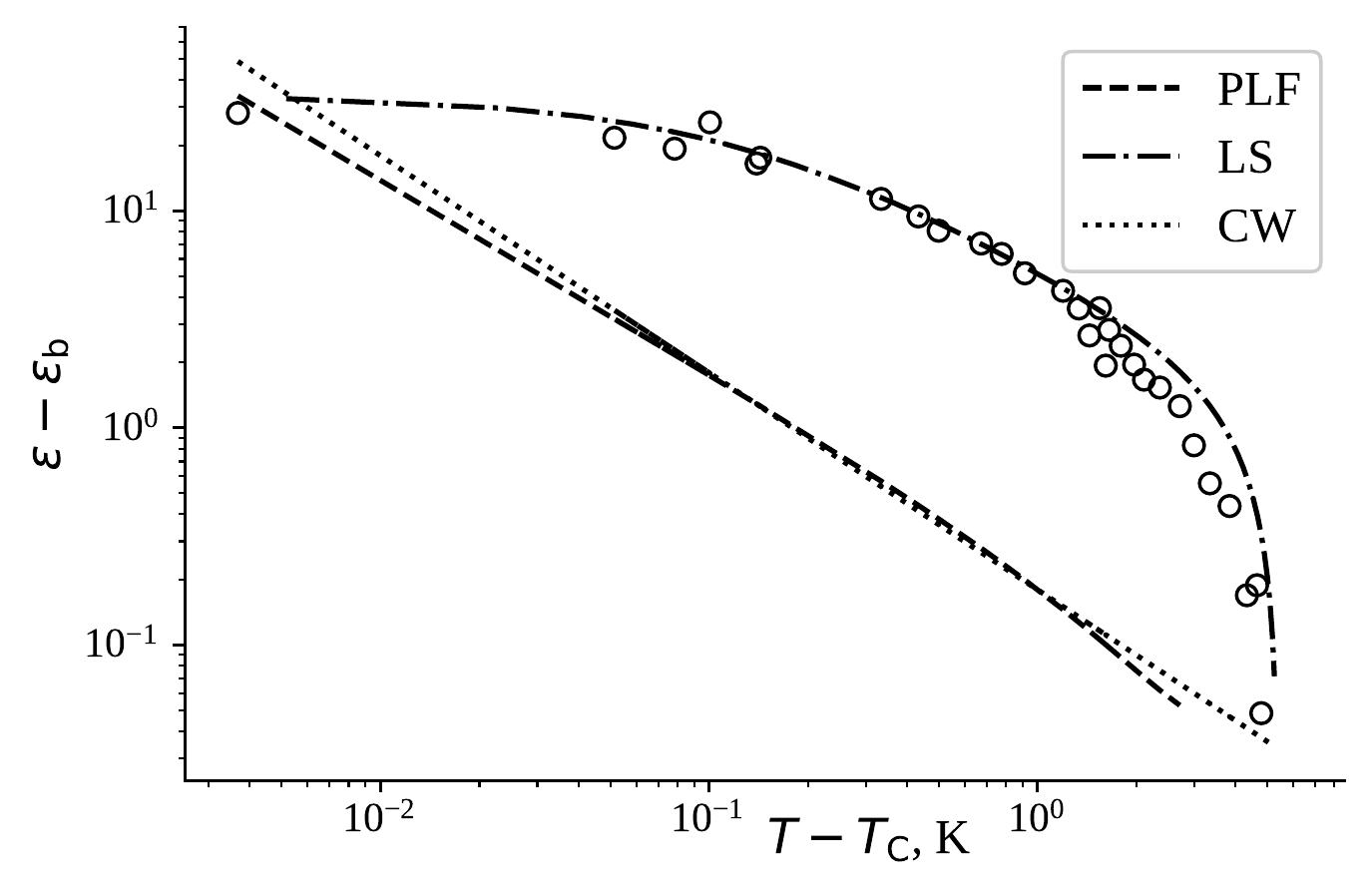} \\ b}
	\end{minipage}
	\caption{Temperature dependences of reciprocal dielectric permittivity (a) and log-log plots of dielectric permittivity (b). Circles correspond to experimental data and lines correspond to different theoretical models: Curie-Weiss (CW), LS and PLF (see the legend). In the both cases, the background term $\varepsilon_b$ is extracted from the experimental data.}
	\label{fig:recip_log_comp}
\end{figure}

Now, we wish to clarify  more scrupulously why, contrary to the model (3), the models (1), (2) and~(4) agree worse with the experimental results for the AFB crystals.
For this purpose, we display both the experimental and theoretical data $\varepsilon(T)$ either in the `Curie-Weiss' coordinates $(\varepsilon - \varepsilon_{\textrm{b}})^{-1}$ vs. $(T - T_{\textrm{C}})$ (see figure~\ref{fig:recip_log_comp}a) or on the double logarithmic scale $\log(\varepsilon - \varepsilon_{\textrm{b}})$ vs. $\log(T - T_{\textrm{C}})$ (see figure~\ref{fig:recip_log_comp}b).
To prevent the overloading of these figures, we do not show the data obtained with the generalized power law (\ref{genpower}).
The results for this model differ only by insignificant details from those illustrated in figure~\ref{fig:recip_log_comp}a and figure~\ref{fig:recip_log_comp}b for the models (1) and (4).

Although the data of figure~\ref{fig:recip_log_comp} can hardly be used for a thorough quantitative interpretation (see the discussion in section \ref{sec:exp_models} and \cite{goldsteinmorrisyen2004, bauke2007, perline2005}), it illustrates well the main practical tendencies for the above models.
The Curie-Weiss law for the dielectric permittivity implies a straight line in the $(\varepsilon - \varepsilon_{\textrm{b}})^{-1}$ vs. $(T - T_{\textrm{C}})$ coordinates and a straight line with the slope $-1$ on the log-log scale (see the dot lines in figure~\ref{fig:recip_log_comp}a and figure~\ref{fig:recip_log_comp}b).
A close examination of the behavior of theoretical PLF function \ref{plfmodel} involving the elliptical integrals testifies that there exists a temperature region where the model (4) can be approximately reduced to the inverse power law, i.e., the Curie-Weiss relation.
Namely, this is an intermediate region above the PT temperature $T_{\textrm{C}}$ given by $T - T_{\textrm{C}} \approx$ 10$^{-1}$--10$^0$~K (see a dashed line in figure~\ref{fig:recip_log_comp}a and, especially, in figure~\ref{fig:recip_log_comp}b).
It corresponds to moderately large reduced temperatures $\tau$ ($\tau \approx$ 0.82--0.98).
Note that formula~(\ref{plfmodel}) was used in the work \cite{levstikprelovsekfilipiczeks1982} for interpretation of the dielectric properties of incommensurate Rb$_2$ZnCl$_4$ crystals, and the authors \cite{levstikprelovsekfilipiczeks1982} actually confirmed that, in the region of intermediate relative temperatures $(T - T_\textrm{C})$, the relation (\ref{plfmodel}) yields the results very close to the Curie-Weiss law.

At the temperatures more distant from $T_{\textrm{C}}$ (i.e., in the region defined by the inequality $T - T_{\textrm{C}} > 1$~K, or at $\tau \approx$ 0--0.82), one can observe severe deviations of the PLF model from the inverse power law (see figure~\ref{fig:recip_log_comp}a).
Finally, the predictions of the PLF model become progressively different from those of the Curie-Weiss law also in the region given by $T - T_{\textrm{C}} < 0.1$~K, i.e. at $\tau >$ 0.98 (see figure~\ref{fig:recip_log_comp}b).
Eventually, both the Curie-Weiss and the PLF models do not claim to consider the fluctuation corrections in the critical region since they correspond to the mean-field theory, which is inapplicable in the closest vicinity of the PT points.

Now that we have ascertained the main formal differences among the theoretical models, we are in a position to compare better the experiment with all the theoretical predictions.
As seen from figure~\ref{fig:recip_log_comp}a, the experimental dependence $\varepsilon(T)$ without the background deviates significantly from the Curie-Weiss law and, moreover, from any other inverse power law.
The latter fact is also evident from figure~\ref{fig:recip_log_comp}b, since the slope of the experimental curve on the log-log scale changes continuously.
Of course, the PLF model predicting nearly inverse power law would fail in describing such  data.
On the contrary, the LS model (see dash-dot lines in figure~\ref{fig:recip_log_comp}) is governed by a combination of terms linear in temperature, which enter  both the numerator and the denominator of formula (\ref{lsmodel}).
This mathematical structure provides a gradual change in the slope of the LS curve in both figure~\ref{fig:recip_log_comp}a and figure~\ref{fig:recip_log_comp}b and so satisfactorily describes  the experimental data.

\section{Comparison of LS and PLF models for the dielectric permittivity} 
\label{sec:comparison}

The fact that the LS model describes better the dielectric permittivity of the AFB crystals than the PLF model is unexpected and even counter-intuitive. 
Indeed, the LS approach looks simpler than the PLF model \cite{prelovseklevstikfilipic1983} which was developed later, the authors \cite{prelovseklevstikfilipic1983} may have been familiar with the LS results~\cite{levanyuksannikov1976} and, moreover, they supposed their model to be applicable in a wider temperature region of the incommensurate phase,  compared with the LS model.
To understand better these problems, we  elucidate in brief the main differences in the physical assumptions underlying the two models.
To do that, we outline the main points of derivation of formulae (\ref{lsmodel}) and (\ref{plfmodel}).

The authors \cite{levanyuksannikov1976, prelovseklevstikfilipic1983}  started from the same free-energy expansion in the framework of the mean-field theory. In polar coordinates, it can be written as follows:
\begin{align}
    \Phi = \frac{\alpha}{2}\rho^2+\frac{\beta_1}{4}\rho^4+\frac{\beta_2}{4} & \rho^4\cos{4\varphi} -\sigma\rho^2\frac{\rd \varphi}{\rd z}+\frac{\delta}{2}\bigg[\bigg(\frac{\rd \rho}{\rd z}\bigg)^2+\rho^2\bigg(\frac{\rd \varphi}{\rd z}\bigg)^2\bigg] \nonumber\\ \
    &- E P + \frac{\kappa}{2} P^2 + a \rho^2 P \cos{2\varphi}.
\label{potential}
\end{align}
Here, $\alpha = \alpha^\prime (T - \theta)$, $\rho$ and $\varphi$ are respectively the amplitude and the phase of the order parameter, $E$ is the electric field, $P$ is the polarization, $\theta$ is the temperature point of structural instability, and $\alpha^\prime$, $\beta_1$, $\beta_2$, $\sigma$, $\delta$, $\kappa$ and $a$ denote temperature-independent constants (see also section \ref{sec:exp_models}).
The most fundamental fact is the availability in formula (\ref{potential}) of a Lifshitz invariant proportional to $\sigma$, which is symmetry-allowed for incommensurately modulated media.
Due to nonzero $\sigma$ term in (\ref{potential}), the initial high-temperature phase does not lose its stability under the condition $\alpha = 0$ (i.e., at $T = \theta$).
This occurs with respect to inhomogeneous displacements at some other value $\alpha = \alpha_0 = \sigma^2 / \delta$ corresponding to a higher paraelectric--incommensurate temperature $T_{\textrm{i}}$.

Note that the $\beta_2$ term in formula (\ref{potential}), which is associated with spatial anisotropy of the order parameter, notably complicates the problem of finding a steady-state solution for the free energy.
Since the above approach is focused first of all on some (small but not too small) vicinity of $T_{\textrm{i}}$, as is always the case with the mean-field theory, strongly anisotropic higher-order terms in the order parameter are omitted in (\ref{potential}).
They drive a lock-in commensurate PT at $T_{\textrm{C}}$ rather than the incommensurate PT at $T_{\textrm{i}}$ (see, e.g., the work \cite{levstikprelovsekfilipiczeks1982}).

Probably, the main difference in the LS and PLF models \cite{levanyuksannikov1976, prelovseklevstikfilipic1983} is the approaches adopted by the authors to solve the system of Lagrange--Euler equations
\begin{equation} \label{le_rho_eq}
    \frac{\partial\Phi}{\partial\rho} - \frac{\rd}{\rd z}\frac{\partial\Phi}{\partial\rho^\prime} = 0,
\end{equation}
\begin{equation} \label{le_phi_eq}
    \frac{\partial\Phi}{\partial\varphi} - \frac{\rd}{\rd z}\frac{\partial\Phi}{\partial\varphi^\prime} = 0,
\end{equation}
where $\rho^\prime = \frac{\rd\rho}{\rd z}$ and $\varphi^\prime = \frac{\rd \varphi}{\rd z}$.
This system of equations allows one to find the values of $\rho$ and $\varphi$ corresponding to stationary state.
Following the work \cite{ishibashi1980}, PLF use a constant-amplitude approximation, according to which the amplitude of the order parameter does not depend on the coordinates [$\rho(z) = \rho_0$]:
\begin{equation}\label{plf_amp}
    \rho_0^2 = \frac{\alpha^\prime}{\beta_1}(T_{\textrm{i}} - T).
\end{equation}
This enables the authors \cite{prelovseklevstikfilipic1983} to come to something like a time-independent sine-Gordon equation,
\begin{equation} \label{plf_phi_eq}
    \frac{\rd^2\varphi}{\rd z^2}=\frac{\beta_2}{4\delta}\rho_0^2\sin{4\varphi}.
\end{equation}
The explicit solution of this equation is found for the temperature behavior of the phase $\varphi$ (see also~\cite{sannikov1981, sannikov1982, sannikov1985}:

\begin{equation} \label{plf_phi}
    2\varphi=\textrm{am}(2qz, \epsilon),
\end{equation}
with  $q^2 = 2\beta_2\rho_0^2/(\delta \epsilon^2)$,  $\epsilon^2 = 2\beta_2\rho_0^4/(C + \beta_2\rho_0^4)$ and $C$ being an integration constant.
Here, $\textrm{am}(2qz, \epsilon)$ represents the Jacobian elliptic function with the modulus $\epsilon$ ($0 \leqslant \epsilon \leqslant 1$). 
Such an approach would imply a full-scale consideration of anisotropy $\beta_2$.

LS treat the same problem in another manner.
They completely neglect  the anisotropy ($\beta_2 = 0$) in the initial stage so that the equation (\ref{plf_phi_eq}) is simplified to $\rd^2 \varphi / \rd z^2 = 0$.
This yields a standard formula for the plane-wave region of incommensurate phase \cite{kushnir1997}:
 \begin{equation} \label{ls_phi}
    \varphi = k_0 z,
\end{equation} 
with the wave vector $k_0 = |\sigma|/\delta$.
On the other hand, in fact LS also start from the constant-amplitude approximation (\ref{plf_amp}) under the condition $\beta_2 = 0$.
As a result, their approach seems to be notably simpler than that of PLF.
However, after that LS take into account higher-order corrections to formula (\ref{ls_phi}) given by a power series in $\beta_2$ (more exactly, in the parameter $\Delta = \frac{\alpha^\prime\delta|\beta_2|}{\sigma^2\beta_1}(T_{\textrm{i}} - T)=\frac{|\beta_2|}{\beta_1}t \ll 1$), the lowest-order of which is proportional to $\Delta^2$ \cite{levanyuksannikov1976}.
This corresponds to what can be termed as a `weak anisotropy approximation', which would eventually affect the final solution for the amplitude, too.

Since the dielectric permittivity $\varepsilon$ is defined as $\varepsilon = \rd P / \rd E$, we obtain
\begin{equation} \label{gen_chi}
    \varepsilon = \frac{1}{\kappa} - \frac{2a \rho}{\kappa}\bigg(\frac{\partial\rho}{\partial E}\cos{2\varphi} - \frac{\partial\varphi}{\partial E} \rho \sin{\varphi}\bigg).
\end{equation}
It is obvious that, unlike the PLF model, the phase in (\ref{ls_phi}) does not depend on thermal changes in the approximation $\Delta = 0$.
Taking derivatives in (\ref{gen_chi}), we arrive at a temperature-independent expression~$\varepsilon(T)$ which coincides with that obtained for the commensurate phase \cite{levanyuksannikov1976}:
\begin{equation}
    \varepsilon_{\textrm{com}}(T) = \frac{1}{\kappa} + \frac{2 a ^2}{\kappa^2 (\beta_1^\prime - | \beta_2^\prime|)},
\end{equation}
with $\beta_1^\prime$ and $\beta_2^\prime$ being renormalized coefficients ($\beta_1^\prime = \beta_1 - 2 a^2/\kappa$ and $\beta_2^\prime = \beta_2 - 2 a^2/\kappa$).
Having left a zero approximation, one can obtain a more complex expression in some vicinity of the incommensurate PT (at $\Delta \ll 1$) (cf. also with the earlier, less correct formula in the work \cite{levanyuksannikov1976_2}):
\begin{equation} \label{ls_chi}
    \varepsilon_{\textrm{LS}}(T) = \frac{1}{\kappa} + \frac{a^2}{\kappa^2 \beta_1^\prime} t \frac{6 + t}{4 - t}.
\end{equation}
Formula (\ref{ls_chi}) coincides with (\ref{lsmodel}) with the notation $\varepsilon_{\textrm{b}} = 1/\kappa$ and $A = a^2/\beta_1^\prime$.
Finally, formulae (\ref{plf_amp}), (\ref{plf_phi}) and (\ref{gen_chi}) obtained in frames of the PLF model result in
\begin{equation} \label{plf_eps}
    \varepsilon_{\textrm{PLF}}(T) = \frac{1}{\kappa} + \frac{a^2}{\kappa^2 \beta_1^\prime}\bigg(\frac{E(\tau)}{(1 - \tau^2) K(\tau)} - 1\bigg),
\end{equation}
where the substitutions $\varepsilon_{\textrm{b}} = 1/\kappa$ and $c = \kappa\beta_1^\prime/a^2$ lead to formula (\ref{plfmodel}).

Now we are in a position to compare the physical backgrounds of the LS and PLF models for the dielectric properties of AFB. 
At the first glance, the PLF result (\ref{plf_phi}) underlying the formula (\ref{plf_eps}) looks stronger  compared to formula (\ref{ls_phi}) obtained by LS, which is limited to a narrow vicinity of paraelectric--incommensurate PT.
However, as pointed out in the work \cite{sannikov1985}, any phenomenological model like those suggested by LS and PLF \cite{levanyuksannikov1976, prelovseklevstikfilipic1983} is anyway applicable only near the PT point $T_{\textrm{i}}$, where the spatial anisotropy is small enough.
In some sense, the approximations of constant amplitude and weak anisotropy have close applicability regions.
Then, the decision of PLF to maintain the exact solution for the phase $\varphi$ and, at the same time, restrict themselves to the limit $\rho(z) = \rho_0$ can prove to be partly inconsistent, as if someone would exceed the accuracy of a given approximation.
Probably, this is the main reason why the PLF formula is less accurate in describing  the $\varepsilon(T)$ function for the AFB crystals.

On the other hand, the fact that the LS model has turned out to work fairly well in the overall range of incommensurate phase can be explained, at least partly, owing to the following circumstance: in terms of a variable $(T_{\textrm{i}} - T_{\textrm{C}})/T_{\textrm{i}}$ characterizing the temperature width of this phase, the latter is very narrow ($\sim$ 0.03).
Eventually, this factor also degrades a potential advantage of the PLF model associated with consideration of the spatial anisotropy, which would have played a more significant part in a wider temperature region.
In this respect, we suppose that an LS-like model could hardly succeed when describing the dielectric properties of Rb$_2$ZnCl$_4$ where the incommensurate phase is very wide ($(T_{\textrm{i}} - T_{\textrm{C}})/T_{\textrm{i}} \sim$ 0.36) and, moreover, the experimental data in the vicinity of  $T_{\textrm{i}}$ are scarce \cite{levstikprelovsekfilipiczeks1982}.

\section{Potential influence of fluctuations and structural defects} 

It is well known that the incommensurate phases in A$_2$BX$_4$ crystals are highly sensitive to any structural imperfections, e.g., due to pinning of the phase of the order parameter \cite{cummins1990}.
This implies that the dielectric permittivity can manifest some dependence on crystal samples or experimental conditions (heating or cooling run, temperature change rate, etc.).
This poses a question of the potential influence of these phenomena on our data and conclusions.
The next question is associated with the effect of the order-parameter fluctuations on the dielectric data.

As stressed above, both the LS and PLF models represent mean-filed approaches.
The temperature region $\delta T = T - T_{\textrm{C}}$ (or $\delta \tau^\prime$ in terms of a redefined reduced temperature, $\delta \tau^\prime = \delta T / T_{\textrm{C}}$) around the phase-transition point where the fluctuations and the critical phenomena begin to dominate and the Landau theory can no longer be employed is given by a so-called Ginzburg parameter $G$: $\tau^\prime \ll G$ or, at least, $\tau^\prime < G $ --- see, e.g., \cite{patashinskiipokrovsky1979, ivanovetal1990}).
The corresponding results derived by us for AFB with a highly sensitive optical-birefringence technique (see \cite{kushniretal2011}) will be reported elsewhere.
Here, we only state that they yield in $G \approx 0.0026$.
Then, we have the conditions $\delta T \ll 0.5$~K or $\delta T < 0.5$~K.
Inspection of the data in figure~\ref{fig:model_comp} (or, better, in figure~\ref{fig:recip_log_comp}b) testifies that some eight data points correspond to the region 0.5~K above $T_{\textrm{C}}$ and only three data points correspond to the region 0.1~K.
In other words, our study does not  directly address the scaling region and includes, at the most, a region where the mean-filed theory can be applied with small fluctuation corrections.
Then, the order-parameter fluctuations can hardly affect our results.

The next point is concerned with `frozen-in fluctuations', i.e., with structural defects (see \cite{levanyuksigov1988}).
Having no direct facilities for estimating a defect state of our sample, we  rely upon indirect methods.
Namely, it is known that the influence of structural defects can disguise itself as a fluctuation effect in a close vicinity of PT.
Therefore, the defects usually widen the `fluctuation' region, i.e., they contribute additively to the Ginzburg parameter (see \cite{kushnirshopavlokh2008, kushniretal2011}).
This enables us to perform a rough comparison of the structural perfection for different samples of a given crystal: the larger is the Ginzburg number obtained for a crystal sample, the higher is the concentration of its defects.
The following fact is worthwhile in this respect.
When comparing our results with the corresponding data for the other A$_2$BX$_4$  crystals \cite{kushniretal2011}, one observes that the Ginzburg parameter for our AFB crystal ($G \sim$ 0.003) is relatively small (although the same in the order of magnitude).
This indirectly indicates that the structural imperfection typical of our crystal sample is not so high to dominate the temperature dependences of its physical properties. 

Moreover, the defects with heavy concentrations can even `smear' the divergent-like anomalies detected at the PT points.
However, we observe no such situation with our sample, thus confirming again that the effects studied by us are nor defect-driven.
Another similar argument against a significant contribution of the defects into the dielectric behavior of our AFB crystal is as follows.
One of the common consequences of strong influence of structural defects is a decrease in the dielectric peak $\varepsilon_{\textrm{max}}$.
However, our parameter $\varepsilon_{\textrm{max}} \approx 55$ is very close to the average value $\varepsilon^{\textrm{avg}}_{\textrm{max}} \approx 57$ found from \cite{strukovetal1973, gesiozawa1974, levanyuksannikov1976, strukovarutyunovauesu1982, prelovseklevstikfilipic1983, jakubasczpala1984, hoshinoetal1958} at comparable electric frequencies.
This is another evidence that the structural defects should play only a secondary role in the dielectric behavior of our crystal sample.

We would also like to emphasize that, in some other terms, our main conclusion is that the temperature anomaly of the dielectric permittivity in the incommensurate phase of improper ferroelectric AFB near the $T_{\textrm{C}}$ point is `slower' than that predicted by the inverse power law (see, e.g., a gradual decrease in the slope --- i.e., the power-law `exponent' --- with approaching $T_{\textrm{C}}$, which is seen in the double logarithmic-scale plot in figure~\ref{fig:recip_log_comp}b), although this law is a common regularity known in the theory of PTs.
A (very loose) analogy with the situation occurring in proper uniaxial ferroelectrics can be mentioned: therein, the leading temperature-dependent terms are also `slower' than those given by the inverse power law, being described by logarithmic corrections.
However, there is still no theory predicting such a `slow-down' in the dielectric divergence near the PT point as a result of structural defects.

Finally, an important question arises in view of a potential effect of structural defects on the dielectric properties of the AFB crystals: is the LS model universally better than the PLF model --- or some experimental data could be found in the literature which prefer the latter model?
Since we cannot rule out completely the sample dependence of the dielectric permittivity, it would be difficult to expect a straightforward answer.
However, this situation seems to be quite unlikely because both of the LS and PLF models refer to defect-free crystals.
Then, the application of these models to essentially imperfect crystal samples might have rather resulted in the failure of both models than in changing the balance of their efficiencies \cite{horonkushnirstadnykkashuba2020}.
\vspace{-3mm}

\section{Conclusions} 

We have studied the dielectric properties of improper ferroelectric AFB crystals in their paraelectric, incommensurately modulated and commensurate ferroelectric phases.
Similar to the previous experimental studies, the dielectric permittivity of AFB is not affected by the incommensurate PT at $T_{\textrm{i}}$ but reveals a weak peak at the commensurate PT point $T_{\textrm{C}}$.
The experimental results for the incommensurate phase of AFB are compared with the data following from the four phenomenological theories: the Curie-Weiss and generalized Curie-Weiss laws and the LS and PLF models \cite{levanyuksannikov1976, prelovseklevstikfilipic1983}.
It is ascertained that the PLF model provides the results very similar to those of the inverse power laws given by the Curie-Weiss and generalized Curie-Weiss formulae.
According to the results of rigorous statistical tests, all of these models provide a much worse fit of the experimental data than the LS model.
In addition, the latter model can be efficiently applied within the overall temperature range of the incommensurate phase.

The analysis of the experimental data shows that the temperature slopes of both the reciprocal permittivity with subtracted dielectric background and the permittivity plotted on the double logarithmic scale change continuously with temperature. 
However, any inverse power law would have implied a constant slope.
This is a formal reason why the models (1), (2) and (4) fail in describing the experimental results.
On the contrary, the temperature dependence of the permittivity in frames of the LS model (3) is governed by a combination of terms linear in temperature, including a divergent term in the denominator.
Then, the peak at the PT point $T_{\textrm{C}}$ is `damped' by the temperature-dependent terms in the numerator.
This mathematical structure provides a necessary change in the slope and so appropriately describes  the experimental data.

In order to compare different phenomenological models more in detail, the main physical hypotheses underlying the LS and PLF approaches are elucidated.
In particular, it is stressed that the LS model is based upon the approximation of weak spatial anisotropy of the order parameter and small corrections to the approximation of constant amplitude of the order parameter.
These approximations are fully justified only within the plane-wave region of the incommensurate phase.
On the other hand, the PLF model employs the constant-amplitude approximation and finds an exact solution for the phase of the order parameter, thus not relying on the assumption of weak anisotropy.
However, the two approximations partly contradict each other, which may be the reason of a lower efficiency of the PLF model,  compared to the LS model.
Most likely, the LS model remains applicable within the entire incommensurate phase in AFB due to a very narrow temperature range of the latter.
This fact also undermines a potential advantage of the PLF approach, i.e., its applicability outside the plane-wave region, when it is applied to the incommensurate crystals like AFB.

Possible contributions of the structural defects and the critical fluctuations into the $\varepsilon(T)$ function of our AFB crystals are discussed.
It is shown that the influence of the defects can hardly be decisive, while the fluctuations typical of a very close vicinity of the PT point are out of the scope of our study and so cannot affect its main conclusions.

\section*{Acknowledgements}

This study has been supported by the Ministry of Education and Science of Ukraine (the Project \#0120U102320).


\ukrainianpart

\title{Температурна залежність діелектричної проникності в несумірно модульованій фазі фторберилату амонію}

\author{Б. І. Горон\refaddr{label1, label2}, О. С. Кушнір\refaddr{label2}, П. А. Щепанський\refaddr{label1}, В. Й. Стадник\refaddr{label1}}
\addresses{
	\addr{label1} Кафедра загальної фізики, Львівський національний університет імені Івана Франка,\\ вул. Драгоманова, 23, 79005 Львів, Україна
	\addr{label2} Кафедра оптоелектроніки та інформаційних технологій, Львівський національний університет\\ імені Івана Франка, вул. ген. Тарнавського, 107, 79013 Львів, Україна}

\makeukrtitle
\begin{abstract}
	\tolerance=3000%
	Досліджено температурну залежність діелектричної проникності вздовж полярної осі сегнетоелектричного кристала фторберилату амонію (ФБА) в околі точок його фазових переходів.
	Експериментальні дані для несумірно модульованої фази ФБА порівняно з передбаченнями феноменологічних моделей, відомих з літератури: закону Кюрі–Вейса (КВ), узагальненого закону Кюрі–Вейса (УКВ), а також моделей Леванюка і Саннікова (ЛС) і Преловшека, Левстіка та Філіпіча (ПЛФ), запропонованих для невласних сегнетоелектриків.
	Показано, що підхід ЛС краще описує температурну поведінку діелектричної проникності для кристала ФБА, ніж моделі КВ, УКВ і ПЛФ.
	З’ясовано основні фізичні причини такої ситуації.
\end{abstract}

\end{document}